\documentstyle [12pt,epsf]{article}
  \large \normalsize

\begin{document}

\begin{center}
\vspace*{1.0cm}

{\Large\bf Optimal Control of Quantum Dynamics : A New Theoretical Approach}

\vspace{1.0cm}

Bijoy K Dey

\vspace*{0.2cm}
{\small \sl Department of Chemistry, University of Toronto, Toronto, Ontario, Canada M5S 3H6\\email:bdey@tikva.chem.utoronto.ca}
\end{center}

\newpage

\begin{center}

\vspace*{0.7cm}

{\bf Abstract}
\end{center}

~~~~~~~~~~~~A new theoretical formalism for the optimal quantum control has been presented. The approach stems from the consideration of describing the time-dependent quantum systems in terms of the real physical observables, viz., the probability density $\rho (x,t)$ and the quantum current j(x,t) which is well documented in the Bohm's hydrodynamical formulation of quantum mechanics. The approach has been applied for manipulating the vibrational motion of HBr in its ground electronic state under an external electric field.

\vspace*{1.0cm}

\newpage
\section{Introduction}
~~~~~~~~~~~~Manipulating the outcome of a chemical dynamics by properly tailoring an external field has been one of the active field of research in the recent days [1-14]. Problems where the external field is an electromagnetic field have received the most attention [2-14], although other applications may arise as well. Theoretically, there are two basic paradigms for such control : a static control scheme [13,14] and a dynamic control scheme [2-12]. In the static scheme [13,14] one uses two or more cw light fields(optical coherence) and the superposition of two or more eigenstates(molecular coherence) to cause interference between different plausible pathways to a final quantum state, and the outcome is controlled by tailoring different parameters of the optical and molecular coherences. Whereas the dynamic scheme [2-12] creates non-stationary states of one's choice, by optimally designing the electric field. This comes under the domain of the optimal control theory [15], a mathematical tool commonly used in the engineering fields. A basic difficulty in attaining the control designs is the computational effort called for in solving the time-dependent Schroedinger equation, often repeatedly in an iterative fashion over an extended spatial region.

~~~~~~~~~~~~In this paper, we introduce a new formulation aiming at reducing the effort for the quantum optimal control(QOC). Our recent work [16,17] has shown that the Bohmian quantum hydrodynamics(BQH) is capable of being much more efficient than the conventional method(e.g., FFT propagation) and this should carry over to the optimal control task. This paper will show how the BQH can be utilized in the QOC. The formulation is based on the hydrodynamic description of the quantum mechanics emerging mainly from the work of David Bohm [18,19] where the dynamics is described by two equations, viz., the equation of motion for the probability density, $\rho(r,t)$ and that for the quantum current, j(r,t) which are defined as $\rho(r,t)=\Psi^*(r,t)\Psi(r,t)$ and $j(r,t)=\frac{1}{2}\frac{\hbar}{m}Im[\Psi^*\nabla\Psi-\Psi\nabla\Psi^*]$, $\Psi$ being the complex wave function in the time dependent Schroedinger equation(TDSE) and Im refers to the imaginary value. Thus one by-passes the explicit use of the time dependent Schroedinger equation(TDSE) and hence the typically oscillatory nature of the complex wave finction. This seems benificiary at the first place because (i) one deals with the real quantum mechanical variables, and (ii) density and quantum current posses a smooth spatial variation as opposed to the wave function. Recent illustrations [16,17] have demonstrated the smooth spatial and temporal nature of the variables and the ability to discretize them on a relatively small number of grid points. In pursuing the paper we maintain the following layout. In section 2 we give a brief account of the BQH. In section 3 we provide the QOC formulation based on the BQH. In section 4 we apply the method for manipulating the vibrational motion of HBr molecule in its ground electronic state. Section 5 concludes the paper. 

\section{Bhomian Quantum Hydrodynamics}
~~~~~~~~~~~~~Despite its extraordinary success, quantum mechanics has since its inception some seventy years ago, been plagued by conceptual difficulties. According to orthodox quantum theory, the complete description of a system of particles is provided by its wave function $\Psi$ which obeys the time-dependent Schr\"{o}dinger equation.
\begin{eqnarray}
i\hbar \frac{\partial \Psi (q,t)}{\partial t}=H\psi (q,t)
\end{eqnarray}
~~~~~~~~~~~~According to Bohm [18], the complete description of a quantum system is provided by its wave function $\Psi (q,t)$, $q \in R^3$, and its configuration $Q \in R^3$ where Q is the position of the particle. The wave function, which evolves according to Schr\"{o}dinger's equation(Eq.(1)) choreographs the motion of the particle which evolves according to the equation
\begin{eqnarray}
\frac{dQ}{dt}=\frac{\hbar}{m}\frac{Im(\Psi ^*\nabla \Psi)}{\Psi ^*\Psi}
\end{eqnarray}
where $\nabla =\frac{\partial}{\partial q}$. In the above equation H is the usual nonrelativistic Hamiltonian for spinless particle given as
\begin{eqnarray}
H=-\frac{\hbar ^2}{2m}\nabla ^2 +V
\end{eqnarray}
~~~~~~~~~~~~Equations (1) and (2) give a complete specification of the quantum theory describing the behaviour of any observables or their effects of measurement. Note that Bohm's formulation incorporates Schr\"{o}dinger's equation into a rational theory, describing the motion of particles, merely by adding a single equation, the guiding equation(Eq.(2)). In so doing it provides a precise role for the wave function in sharp contrast with its rather obscure status in orthodox quantum theory. The additional equation(Eq.(2)) emerges in an almost inevitable manner. Bell's preference is to observe that the probability current $j^\Psi$ and the probability density $\rho =\Psi ^*\Psi$ would classically be related by $j=\rho v$ obviously suggests that
\begin{eqnarray}
\frac{dQ}{dt}=v=j/\rho
\end{eqnarray}
~~~~~~~~~~~Bohm, in his seminal hidden-variable paper wrote the wave function $\Psi$ in the polar form $\Psi =R e^{iS/\hbar}$ where S is real and $R\ge 0$, and then rewrote the schr\"{o}dingers's equation in terms of these new variables, obtaining a pair of coupled evolution equations, the continuity equation for $\rho =R^2$ as
\begin{eqnarray}
\frac{\partial \rho}{\partial t}=-\nabla .(\rho v)
\end{eqnarray}
which suggests that $\rho$ be interpreted as a probability density, and a modified Hamilton-Jacobi equation for S,
\begin{eqnarray}
\frac{\partial S}{\partial t}+H(\nabla S,q)+V_q = 0
\end{eqnarray}
where $H=H(p,q)$ is the classical Hamiltonian function corresponding to Eq.(3),
and
\begin{eqnarray}
V_{q}&=&-\frac{\hbar ^2}{2m}\frac{\nabla ^2 R}{R}\\\nonumber
&&=-\frac{\hbar ^2}{2m}\nabla ^2ln\rho ^{1/2}-\frac{\hbar ^2}{2m}(\nabla ln\rho
^{1/2})^2
\end{eqnarray}
~~~~~~~~~~~~Eq.(6) differs from the classical Hamilton-Jacobi equation only by the appearence of an extra term, the quantum potential $V_q$. Similar to the classical Hamilton-Jacobi equation, Bohm defined the quantum particle trajectories, by indentifying $\nabla S$ with $mv$, by
\begin{eqnarray}
\frac{dQ}{dt}=\frac{\nabla S}{m}
\end{eqnarray}
which is equivalent to Eq.(4). This is precisely what would have been obtained
classically if the particles were acted upon by the force generated by quantum potential in addition to the usual forces. Although an interpretation in classical terms is beautufully laid down in the above equations, one should keep in mind that in so doing, the linear Schr\"{o}dinger equation is transformed into a highly nonlinear equations(eqs.(5) and (6)). By taking the gradient on both sides of Eq.(6) we obtain
\begin{eqnarray}
\frac{\partial }{\partial t}\mbox{\boldmath$v$}=-(\mbox{\boldmath$v$}.\nabla )\mbox{\boldmath$v$}-\mbox{\boldmath$v$}\times (\nabla \times \mbox{\boldmath$v$})-\frac{1}{m}\nabla (V+V_q)
\end{eqnarray}
Defining the quantum current as $\mbox{\boldmath$j$}(q,t)=\frac{1}{2}\frac{\hbar}{m}Im[\Psi ^*(q,t)\nabla \Psi (q,t)-\Psi(q,t)\nabla\Psi^*(q,t)]=\rho(q,t)\mbox{\boldmath$v$}(q,t)$ and using the equation $\nabla \times \mbox{\boldmath$v$}=0$ we readily obtain the expression for the motion of the quantum current as

\begin{eqnarray}
\frac{\partial}{\partial t}\mbox{\boldmath$j$}=-\mbox{\boldmath$v$}(\nabla .\mbox{\boldmath$j$})-(\mbox{\boldmath$j$}.\nabla)\mbox{\boldmath$v$}-\frac{\rho}{m}\nabla (V+V_q)
\end{eqnarray}

~~~~~~~~~~~~Eqs.(5), (6), (9) and (10) describe the motion of a quantum particle in the hydrodynamical representation of TDSE. However, the many-particle description of the BQH can be found elsewhere [24]. It may be noted that density alone cannot sufficiently describe a quantum system, one requires both density and the quantum current for the purpose. As is evident, the motion of a quantum particle is governed by the quantum current vector $\mbox{\boldmath$j$}$ unlike the TDSE where the time propagator $e^{iHt}$ has the key role for the particle's motion. The difficulties arising out of the evaluation of the exponential of an operator in more than one dimension is completely bypassed in the HDE. Although the hydrodynamical equations resemble the classical fluid dynamical equations, the quantum identity is prevailed because of the fact that the quantum current evolves with respect to a potential $V_q$ which has no classical analogue [19]. It should be be noted that the term $V_q$ was inherrently present in the expression to stabilize the hydrodynamical approach to the TDSE. The numerical instability in the hydrodynamical approach to the TDSE without the presence of $V_q$ term may be related to the ``shock'' formation in the classical hydrodynamics(cf. Navier Stokes equation) without some fictitious smoothing potential. In the numerical solution we shall work with the equations governing the motion of density(Eq.(5)) and the quantum current(Eq.(10)). The motivations of considering the above equations lie in the fact that (i) density $\rho$ and quantum current J are uniquely defined for a given potential in many-body system whereas the phase S can be multivalued(S=S$\pm $ n$\pi $, n=even) and (ii) they are quantum mechanical observables. The equations 5 and 6 suggest that one can obtain density and the quantum current directly for $t > 0$ provided, the values were known at $t = 0$. Thus, the scheme by-passes the evaluation of the wave function during the the occurrence of the dynamics. However, at t=0, one has to solve the time-independent Schr\"{o}dinger equation for the wave function and calculate $\rho(q,0)$ and $j(q,0)$.

\section{Quantum Optimal Control and Bhomian Quantum Hydrodynamics}

~~~~~~~~~~~Quantum optimal control theory seeks the design of an external field to fulfill a particular objective. This section will provide the rigorous mathematical formulation of the hydrodynamic method to design an optimal time dependent field that drives a quantum wave packet to a desired objective at the target time t=T. For this purpose, consider a general target expectation value defined as $\Theta_T=\int_{0}^{T}\Theta\rho (x,T)dx$, where $\Theta$ is an observable operator and $\rho (x,T)$ is the probability density which obeys the hydrodynamical equations, viz., Eqs.(5) and (10). The goal is to steer $\Theta _T$ as close as possible to a desired value $\Theta ^d$. We define a quadratic cost functional as
\begin{eqnarray}
J_q=\frac{1}{2}\omega _a(\Theta _T-\Theta^d)^2
\end{eqnarray}
Minimization of $J_q$ amounts to the equalization of $\Theta _T$ to $\Theta^d$. However, $\rho $ in the above equation must obey the hydrodynamical equations, viz., Eqs.(5) and (10). Thus, we have to fulfill this constraint whereby we obtain the unconstrained cost functional as 
\begin{eqnarray}
\bar{J}&=&J_q-\int\int\lambda _1(x,t)[\frac{\partial \rho(x,t)}{\partial t}+\frac{\partial j(x,t)}{\partial x}]dx dt\\\nonumber
&&-\int\int\lambda _2(x,t)[\frac{\partial j(x,t)}{\partial t}+\frac{\partial }{\partial x}(\frac{j^2}{\rho })+\rho\frac{\partial }{\partial x}(V+V_q+V_{ext}(t))]dx dt
\end{eqnarray}  
where $V_{ext}(t)$ represents the external potential due to the interaction between the particle and the electric field, $E(t)$ to be designed.

~~~~~~~~~~~~Thus, in the above equations, we have introduced two Lagrange's multipliers $\lambda _1(x,t)$ and $\lambda _2(x,t)$. There exists another constraint involving the total energy in the field which must be imposed on the optimization procedure. This constraint takes the form
\begin{eqnarray}
\frac{1}{2}\omega _e[\int_{0}^{T} E^2(t) dt -E_p]=0
\end{eqnarray}
where $E_p$ is the energy of the pulse and E(t) the field to be designed. The parameters $\omega _a$ and $\omega _e$ are the positive weights balancing the significance of the two terms viz., $J_q$ and $J_e=\frac{1}{2}\omega _e\int_{0}^{T} E^2(t) dt$ respectively. The term $J_e=\frac{1}{2}\omega _e\int_{0}^{T} E^2(t) dt$ represents the penalty due to the fluence of the external field. So the full unconstrained cost functional takes the form :
\begin{eqnarray}
\bar{J}&=&J_q-\int\int\lambda _1(x,t)[\frac{\partial \rho(x,t)}{\partial t}+\frac{\partial j(x,t)}{\partial x}]dx dt\\\nonumber
&&-\int\int\lambda _2(x,t)[\frac{\partial j(x,t)}{\partial t}+\frac{\partial }{\partial x}(\frac{j^2}{\rho })+\rho\frac{\partial }{\partial x}(V+V_q+V_{ext}(t))]dx dt\\\nonumber
&&+\frac{1}{2}\omega _e[\int_{0}^{T} E^2(t) dt -E_p]
\end{eqnarray}
In this equation $\bar{J}$ is seen to be a functional of five functions, viz., $\rho (x,t)$, j(x,t), $\lambda _1(x,t)$, $\lambda _2(x,t)$ and E(t), all of which are real, unlike in the conventional method [6-8]. In the above equations capital J is for cost functional and small j is for quantum current. The total variation of $\bar{J}$ can be written as
\begin{eqnarray}
\delta \bar{J}&=&\int\int\frac{\delta \bar{J}}{\delta \rho (x,t)}\delta \rho (x,t)dx dt+\int\int\frac{\delta \bar{J}}{\delta j(x,t)}\delta j(x,t)dx dt\\\nonumber
&&+\int\int\frac{\delta \bar{J}}{\delta \lambda _1(x,t)}\delta \lambda _1(x,t)dx dt+\int\int\frac{\delta \bar{J}}{\delta \lambda _2(x,t)}\delta \lambda _2(x,t)dx dt+\int\int\frac{\delta \bar{J}}{\delta E(t)}\delta E(t)dx dt
\end{eqnarray}
For any optimal solution $\delta \bar{J}=0$, which gives
\begin{eqnarray}
\frac{\delta \bar{J}}{\delta \rho (x,t)}=\frac{\delta \bar{J}}{\delta j(x,t)}=\frac{\delta \bar{J}}{\delta \lambda _1(x,t)}=\frac{\delta \bar{J}}{\delta \lambda _2(x,t)}=\frac{\delta \bar{J}}{\delta E(t)}=0
\end{eqnarray}
We have provided in Appendix-A the full expression for $\delta \bar{J}$. Comparing Eq.(16) with Eq.(A.14)(see Appendix-A) we obtain 
\begin{eqnarray}
\frac{\delta \bar{J}}{\delta\lambda _1(x,t)}=-\frac{\partial \rho}{\partial t}-\frac{\partial j}{\partial x}=0
\end{eqnarray}
\begin{eqnarray}
\frac{\delta \bar{J}}{\delta \lambda _2(x,t)}=-\frac{\partial j}{\partial t}-\frac{\partial }{\partial x}(\frac{j^2}{\rho})-\rho\frac{\partial }{\partial x}(V+V_q+V_{ext}(t))=0
\end{eqnarray}
\begin{eqnarray}
\frac{\delta \bar{J}}{\delta j(x,t)}=\frac{\partial \lambda _2}{\partial t}+\frac{\partial \lambda _1}{\partial x}+2\frac{\partial }{\partial x}(\lambda _2\frac{j}{\rho})-2\frac{\lambda _2}{\rho }\frac{\partial j}{\partial x}+2\frac{\lambda _2j}{\rho ^2}\frac{\partial \rho}{\partial x}=0
\end{eqnarray}
\begin{eqnarray}
\frac{\delta \bar{J}}{\delta \rho (x,t)}&=&\frac{\partial \lambda _1}{\partial t}+2\frac{\lambda _2j}{\rho ^2}\frac{\partial j}{\partial x}-\frac{\partial }{\partial x}(\lambda _2\frac{j^2}{\rho ^2})-2\frac{\lambda _2j^2}{\rho ^3}\frac{\partial \rho}{\partial x}\\\nonumber
&&-\lambda _2\frac{\partial }{\partial x}(V+V_q+V_{ext}(t))-\frac{1}{2\mu\rho ^{1/2}}\frac{\partial ^2}{\partial x^2}(\frac{1}{\rho ^{1/2}}\frac{\partial }{\partial x}(\lambda _2\rho))\\\nonumber
&&+\frac{1}{4\mu\rho ^{3/2}}\frac{\partial ^2}{\partial x^2}\rho ^{1/2}\frac{\partial }{\partial x}(\lambda _2\rho )=0
\end{eqnarray}
\begin{eqnarray}
\frac{\delta \bar{J}}{\delta \rho (x,T)}=\omega _a[\Theta _T-\Theta^d]x-\lambda _1(x,T)=0
\end{eqnarray}
\begin{eqnarray}
\frac{\delta \bar{J}}{\delta j(x,T)}=-\lambda _2(x,T)=0
\end{eqnarray}
\begin{eqnarray}
\frac{\delta \bar{J}}{\delta E(t)}=\int\lambda _2(x,t)\rho (x,t)\frac{\partial }{\partial x}\mu (x)dx +\omega _eE(t)=0
\end{eqnarray}
Eq.(19) and (20) can be rewritten in a simple form as
\begin{eqnarray}
\frac{\partial \lambda _2}{\partial t}+\frac{\partial}{\partial x}(\lambda _2v_{\lambda})+S_1[\rho ,j,\lambda _2]=0
\end{eqnarray}
and
\begin{eqnarray}
\frac{\partial \lambda _1}{\partial t}+\frac{\partial}{\partial x}(\lambda _1v_{\lambda})-\lambda _2\frac{\partial}{\partial x}(V+V_q(\lambda _2)+V_{ext})+S_2[\rho ,j,\lambda _2]=0
\end{eqnarray}
where
\begin{eqnarray}
S_1=-2\frac{\lambda _2}{\rho }\frac{\partial j}{\partial x}
\end{eqnarray}
and
\begin{eqnarray}
S_2&=&-\lambda _2\frac{\partial}{\partial x}(V_q(\rho)-V_q(\lambda _2))-\frac{j^2}{\rho ^2}\frac{\partial \lambda _2}{\partial x}\\\nonumber
&&-\frac{1}{4\rho ^{1/2}}\frac{\partial ^2}{\partial x^2}[\frac{1}{\rho ^{1/2}}\frac{\partial}{\partial x}(\lambda _2\rho )]\\\nonumber
&&+\frac{1}{4\rho ^{3/2}}\frac{\partial ^2}{\partial x^2}\rho ^{1/2}\frac{\partial}{\partial x}(\lambda _2\rho )
\end{eqnarray}

~~~~~~~~~~~~Note that the above expression for $\Theta _T$ restricts the operator $\Theta$ being only a multiplicative operator, for example, the distant $\hat{x}$ which we have used in the subsequent numerical calculations. However, other forms of operator can also be considered in the BQH-QOC formulation with the different constraint expressions, e.g., if $\Theta$ is the momentum operator$(\hat{p})$ we would require the constraint equations (5) and (9) since $p_T=m\int_0^T\rho(x,T)\nabla S(x,T) dx$.

~~~~~~~~~~~~The equations for $\lambda _1$ and $\lambda _2$ ressemble to that of $\rho $ and j with only difference being the extra source terms $S_1$ and $S_2$. The source terms depend on $\rho $ and j. $v_{\lambda}$ in the above equations is the velocity associated with the Lagrange's multiplier and is given as $v_{\lambda}=\frac{\lambda _1}{\lambda _2}$ and $V_q(\lambda _2)$ is given by $V_q(\lambda _2)=-\frac{\hbar ^2}{2\mu}\frac{\nabla ^2\lambda _2^{1/2}}{\lambda _2^{1/2}}$. Notice that Eqs.(17) and (18) are the equations of motion for the probability density and the quantum current density respectively obtained in section 2. Whereas Eqs.(24) and (25) are the equations of motion for the two Lagrange's multipliers $\lambda _2$ and $\lambda _1$ respectively. It should be noted that in obtaining the above equations(see Appendix-A) we have assumed no variation on either $\rho (x,0)$ or j(x,0). Thus, we start from an initial(t=0) $\rho (x,0)$ and j(x,0) to solve Eqs.(17) and (18) for $\rho (x,t)$ and j(x,t) respectively. Equations (24) and (25) can be solved for $\lambda _1(x,t)$ and $\lambda _2(x,t)$ provided a starting value $\lambda _1(x,t_s)$ and $\lambda _2(x,t_s)$ were known. These have been obtained from Eqs.(21) and (22) respectively as 
\begin{eqnarray}
\lambda _1(x,t_s)=\omega _a[\Theta _T-\Theta^d]x \ and \  \lambda _2(x,t_s)=0
\end{eqnarray}
 where $t_s=T$, the final time. Thus, one has to perform backward propagation for solving both the equations of motion involving $\lambda _1(x,t)$ and $\lambda _2(x,t)$. Having calculated $\rho (x,t)$, j(x,t), $\lambda _1(x,t)$ and $\lambda _2(x,t)$ as described above, one has to carry out an optimization of the quadratic cost functional(Eq.(11)) with respect to the electric field E(t) which, according to Eq.(23), takes the form
\begin{eqnarray}
E(t)=-\frac{1}{\omega _e}\int\lambda _2(x,t)\rho (x,t)\frac{\partial }{\partial x}\mu (x)dx
\end{eqnarray}
This constitute the details of the BQH-QOC method. 

\section*{4 Application to HBr Molecule} 

~~~~~~~~~~~~We have said in the preceeding section that we needed the initial density $\rho (x,0)$ and the quantum current j(x,0) in the presnt method. These have been evaluated by solving the time independent Schroedinger equation for HBr molecule in the $^1\sum^+$ state where the the potential energy is assumed Morse type of the form [20]

\begin{eqnarray}
V=D_e(1-exp(-\beta(x-x_e)))^2
\end{eqnarray}
where $\beta=\omega _e(\frac{\mu}{2D_e})^{1/2}$, $D_e=\frac{\omega _e^2}{4\omega _e x_e}$ with $\omega _e=2648.975 cm^{-1}$, $\omega _ex_e=45.217 cm^{-1}$, $x_e=1.41443$ angstrom and $\mu$ being the reduced mass of HBr.
 
~~~~~~~~~~~~Having obtained $\rho (x,0)$ and j(x,0) we carry out the control by the present method. Followings are the necessary steps for the computer implementation of the present method :

{\bf A. Present Method :}
\newcounter{fig}
\begin{list}{\bfseries\upshape Step \arabic{fig}:}
{\usecounter{fig}
\setlength{\labelwidth}{2cm}\setlength{\leftmargin}{2.6cm}
\setlength{\labelsep}{0.5cm}\setlength{\rightmargin}{1cm}
\setlength{\parsep}{0.5ex plus0.2ex minus 0.1ex}
\setlength{\itemsep}{0ex plus0.2ex} \slshape}
\item Make an initial guess for the electric field E(t), which is zero in our calculation.
\item Solve the coupled equations, viz., Eq.(17) and (18) for $\rho (x,t)$ and j(x,t) respectively starting from $\rho (x,0)$ and j(x,0). The solution is done by using the Flux-corrected transport(FCT) algorithm [21] modified by us for the purpose of solving the quantum hydrodynamical equations [16,17]. In doing so, we adopt the Eulerian scheme
\item Evaluate the final values for $\lambda _1(x,T)$ and $\lambda _2(x,T)$ given by Eq.(28).
\item Use $\lambda _1(x,T)$ and $\lambda _2(x,T)$ for solving Eqs.(24) and (25) for $\lambda _1(x,t)$ and $\lambda _2(x,t)$ respectively. This is done by backward propagation, by putting dt=-dt(see ref.16). We follow the same method as in step 2 for solving these equations. It should be noted that Eqs.(24) and (25) have source terms which depend on $\rho (x,t)$ and j(x,t) calculated in step 2. 
\item Calculate the quadratic cost functional given by Eq.(11).
\item Optimize the function in Eq.(11) with respect to the electric field, E(t) given by Eq. (29). Here we use the conjugate direction search method [22] for the optimization.
\item Iterate step 2 to step 6 until a convergence criterion is satisfied.
\end{list}
~~~~~~~~~~~~The external potential is of the form $V_{ext}(x,t)=-\mu (x)E(t)$, where $\mu (x)$ is the dipole function for HBr and is given by [23] $\mu (x)=A_0+A_1(x-x_e)+A_2(x-x_e)^2$ where $A_0=0.788$, $A_1=0.315$ and $A_2=0.575$. In our calculation the range of spatial dimension is $0\le x \le 12$ a.u., that of time is $0\le t \le 2000$ a.u. Total number of spatial mesh points is 60 which gives $\Delta x=0.2$ a.u. Similarly, total number of time steps is 2000, which corresponds to $\Delta t=1.0$ a.u. $\omega _e$ in Eq.(27) is taken as 0.5, and $\omega _a$ as 1000. The target operator is $\Theta =x$ and $\Theta^d=$ 3.0 a.u. and 3.5 a.u.

~~~~~~~~~~~~Figure 1 shows the electric fields corresponding to two different values of $\Theta ^d$ viz., 3.0(solid lines) and 3.5(dotted lines). These pulses excite several vibrational states(not shown here) mainly by a sequence of single quantum transitions. The peak value of the field is $\approx 0.08 a.u.$(corresponding intensity is $\approx 10^{14} W cm^{-2}$) for $\Theta^d=3.5 a.u.$ and $\approx 0.02 a.u.$(corresponding intensity is $\approx 10^{13} W cm^{-2}$) for $\Theta^d=3.0 a.u.$. The detail characterization of the optimal field can however, be made by Fourier transforming the field. Fig.2 shows the average distance $<x>$ as a function of time. Notice the desired control of $<x>=3.0$ and 3.5 a.u. at T=2000 a.u. is obtained through the oscillatory motion of the packed induced by the optimal electric pulse(Fig.1). Figure 3 shows the initial and final densities for the two values of $\Theta ^d$. The packet is distorted in shape while approximately retaining its original variance during the evolution. During the optimization process the total integrated probability density remained at unity up to a deviation of $10^{-7}$. The number of iterations in the optimization to achieve the results is 5 and it takes only 3 minutes(real) on a IRIX IP30 machine with R4400 6.0 CPU. As a test for the acceptability of the present method we have carried on the following experiment : The electric fields(Fig.1) so obtained have been pluged into the TDSE and then solved for the wave function. The results for the density and the expectation value of $<x>$ resemble accurately to that given in Fig.2.
\section*{5. Conclusion}

~~~~~~~~~~~~In the present paper we have presented a new scheme for carrying out the optimal design based on BQH. We have derived the control equations to obtain a time dependent external field with an illustration for the manipulation of the vibrational motion of HBr molecule in the $^1\sum^+$ state. The working dynamical variables in the BQH , viz., $\rho$(Fig.3), j, $\lambda _1$ and $\lambda_2$ are relatively slowly varying spatial functions(Fig. 4) compared to the wave function(fig.4, curve a) which apparently enhances the efficiency and the numerical saving of the BQH-QOC method for controlling dynamics. 

~~~~~~~~~~~~Although the illustration of our new method has been made in one spatial dimension, the approach is general and is directly extendable to higher dimensions and a wave packet dynamics in four dynamics has already been performed [16] within our method. The use of the alternating direction implicit(ADI) [16,17] in the present method makes the quantum control calculation much easier compared to the conventional method, especially for the multidimensional problem. In the conventional optimal control theory, the role of the complex Lagrange's multiplier is to provide feedback[6] for designing the electric field and guide the dynamics to an acceptable solution. The BQH-QOC method, on the other hand, introduces two such Lagrange's multipliers, $\lambda _1$ and $\lambda _2$ both of which are real variables. The first Lagrange's multiplier $\lambda _1$, which corresponds to the quantum current j(cf. Eq.(18) and (25)) has however, no direct role to provide feedback for designing the electric field(Eq.(29)) and only guides the dynamics in conjunction with the second Lagrange's multiplier $\lambda _2$. It may be worth mentioning that since the quadratic cost functional(Eq.(11)) is a functional of density, the Lagrange's multiplier $\lambda _2$(equivalent to the density $\rho $, cf. Eq.(17) and (24)) enters into the expression for the optimal electric field(Eq. (29)). However, cases where one desires to manipulate the quantum flux(which is directly related to the quantum current j) by constructing a quadratic cost functional dependent on j, the Lagrange's multiplier $\lambda _1$ will appear explicitly into the expression for the external field.

~~~~~~~~~~~~It should be pointed out that the present method could prove hard in cases where the dynamics may lead to the creation of the nodes in the density profile since the quantum potential appearing in the constraint equation blows up in the occurence of such an event. However, such occurence of nodes can be countered by fixing a lower limit to the density of the order of the machine precision. This in other words means that one never encounters an absolute nodal point where the density is exactly zero. Future studies need to explore the other area of control within the BQH-QOC method, for example, controlling the quantum flux.
\section*{Acknowledgement}
~~~~~~~~~~~~We thank Dr.Jair Botina for his help.
\begin{center}
{\bf APPENDIX}
\end{center}
The variation of $\bar{J}$ given by Eq.(14) has to be taken with respect to $\rho (x,t)$, j(x,t), $\lambda _1(x,t)$, $\lambda _2(x,t)$ and E(t). Any variation $\delta \rho (x,t)$, $\delta j(x,t)$, $\delta \lambda _1(x,t)$, $\delta \lambda _2(x,t)$ and $\delta E(t)$ will lead to the variation $\delta \bar{J}$ given as
\begin{eqnarray*}
\delta \bar{J}&=&\delta J_q-\int\int [\frac{\partial \rho}{\partial t}+\frac{\partial j}{\partial x}]\delta \lambda _1dxdt\\\nonumber
&&-\int\int [\frac{\partial j}{\partial t}+\frac{\partial}{\partial x}(\frac{j^2}{\rho})+\rho\frac{\partial }{\partial x}(V+V_q+V_{ext})]\delta \lambda _2dxdt \hspace{2.5cm} (A.1)\\\nonumber
&&-\int\int\lambda _1[\frac{\partial }{\partial t}\delta \rho+\frac{\partial }{\partial x}\delta j]dxdt\\\nonumber
&&-\int\int \lambda _2[\frac{\partial }{\partial t}\delta j+\frac{\partial }{\partial x}(\frac{2j}{\rho}\delta j-\frac{j^2}{\rho ^2}\delta \rho)+\delta \rho \frac{\partial }{\partial x}(V+V_{q}+V_{ext})\\\nonumber
&&+\rho\frac{\partial }{\partial x}(\delta V+\delta V_q+\delta V_{ext})]dxdt+\omega _e\int E(t)\delta E(t) dt
\end{eqnarray*}
Now, we have $\delta V=0$ and $\delta V_{ext}$ can be written as $\delta V_{ext}=-\delta (\mu (x)E(t))=-\mu (x)\delta E(t)-E(t)\delta \mu (x)$. Since $\mu (x)$ is kept fixed, we get $\delta V_{ext}=-\mu (x)\delta E(t)$. $J_q$ in the above equation is given by
\begin{eqnarray*}
J_q=\frac{1}{2}\omega _a[\int\rho (x,T)x dx-x_{cm}^d]^2 \hspace{2.5cm} (A.2)
\end{eqnarray*}
Hence
\begin{eqnarray*}
\delta J_q=\omega _a[<x>(T)-x_{cm}^d]\int x\delta \rho (x,T) dx \hspace{2.5cm} (A.3)
\end{eqnarray*}
Substituting Eqs.(A.3) into Eq.(A.1) we obtain
\newpage
\begin{eqnarray*}
\delta \bar{J}&=&\omega _a[<x>(T)-x_{cm}^d]\int x\delta \rho (x,T)dx-\int\int[\frac{\partial \rho}{\partial t}+\frac{\partial j}{\partial x}]\delta \lambda _1 dxdt\\\nonumber
&&-\int\int [\frac{\partial j}{\partial t}+\frac{\partial }{\partial x}(\frac{j^2}{\rho })+\rho \frac{\partial }{\partial x}(V+V_q+V_{ext})]\delta \lambda _2dxdt\\\nonumber
&&-\int\int\lambda _1\frac{\partial }{\partial t}\delta \rho dxdt-\int\int\lambda _1\frac{\partial }{\partial x}\delta j dx dt\\\nonumber
&&-\int\int\lambda _2\frac{\partial }{\partial t}\delta j dxdt-2\int\int\lambda _2\frac{j}{\rho}\frac{\partial }{\partial x}\delta j dxdt \hspace{2.5cm} (A.4)\\\nonumber
&&-2\int\int\lambda _2\frac{1}{\rho}\frac{\partial j}{\partial x}\delta j dxdt +2\int\int\lambda _2\frac{j}{\rho ^2}\frac{\partial \rho}{\partial x}\delta jdxdt\\\nonumber
&&+\int\int\lambda _2\frac{j^2}{\rho ^2}\frac{\partial }{\partial x}\delta \rho dxdt+2\int\int\lambda _2\frac{j}{\rho ^2}\frac{\partial j}{\partial x}\delta \rho dxdt\\\nonumber
&&-2\int\int\lambda _2\frac{j^2}{\rho ^3}\frac{\rho}{\partial x}\delta \rho dxdt-\int\int\lambda _2\frac{\partial }{\partial x}(V+V_q+V_{ext})\delta \rho dxdt\\\nonumber
&&-\int\int\lambda _2\rho \frac{\partial }{\partial x}\delta V_q dxdt+\int\int\lambda _2\rho \frac{\partial }{\partial x}(\mu (x)\delta E(t))dxdt\\\nonumber
&&+\omega _e\int E(t)\delta E(t)dt
\end{eqnarray*}
The 4-th and 6-th terms in the above equation can be simplified by integration by parts as follows 
\begin{eqnarray*}
\int\int\lambda _1\frac{\partial }{\partial t}\delta \rho dxdt&=&\int\lambda _1(x,T)\delta \rho (x,T)dx-\int\lambda _1(x,0)\delta \rho (x,0)dx\\\nonumber
&&-\int\int\frac{\partial \lambda _1}{\partial t}\delta \rho (x,t)dxdt \hspace{2.5cm} (A.5)
\end{eqnarray*}
\begin{eqnarray*}
\int\int\lambda _2\frac{\partial }{\partial t}\delta j dxdt&=&\int\lambda _2(x,T)\delta j(x,T)dx-\int\lambda _2(x,0)\delta j(x,0)dx\\\nonumber
&&-\int\int\frac{\partial \lambda _2}{\partial t}\delta j(x,t)dxdt \hspace{2.5cm} (A.6)
\end{eqnarray*}
Terms 5-th, 7-th and 10-th can similarly be expressed by the integration by bparts as follows 
\begin{eqnarray*}
\int\int\frac{\partial }{\partial x}\delta j dxdt&=&\int\lambda _1(x_r,t)\delta j(x_r,t)dt-\int\lambda _1(x_l,t)\delta j(x_l,t)dt\\\nonumber
&&-\int\int\frac{\partial \lambda _1}{\partial x}\delta j(x,t)dxdt \hspace{2.5cm} (A.7)
\end{eqnarray*}
\begin{eqnarray*}
\int\int\lambda _2\frac{j}{\rho }\frac{\partial }{\partial x}\delta j dxdt&=&\int\frac{\lambda _2(x_r,t)j(x_r,t)}{\rho (x_r,t)}\delta j(x_r,t)dt-\int\frac{\lambda _2(x_l,t)j(x_l,t)}{\rho (x_l,t)}\delta j(x_l,t)dt\\\nonumber
&&-\int\int\frac{\partial }{\partial x}(\lambda \frac{j}{\rho})\delta j dxdt \hspace{2.5cm} (A.8)
\end{eqnarray*}
\begin{eqnarray*}
\int\int\lambda _2\frac{j^2}{\rho ^2}\frac{\partial }{\partial x}\delta \rho dxdt&=&\int\frac{\lambda _2(x_r,t)j^2(x_r,t)}{\rho ^2(x_r,t)}\delta \rho (x_r,t)dt-\int\frac{\lambda _2(x_l,t)j^2(x_l,t)}{\rho ^2(x_l,t)}\delta \rho (x_l,t)dt\\\nonumber
&&-\int\int\frac{\partial }{\partial x}(\lambda _2\frac{j^2}{\rho ^2})\delta \rho dxdt \hspace{2.5cm} (A.9)
\end{eqnarray*}
Term 15-th is
\begin{eqnarray*}
\int\int\lambda _2\rho \frac{\partial }{\partial x}(\mu (x)\delta E(t))dxdt=\int\int\lambda _2\rho \frac{\partial }{\partial x}\mu (x)\delta E(t)dxdt \hspace{2.5cm} (A.10)
\end{eqnarray*}
14-th term involves the variation in $\bar{J}$ due to the change in the quantum potential $\delta V_q$, where $V_q$ is given by $V_q=-\frac{\hbar ^2}{2\mu }\frac{\nabla ^2\rho ^{1/2}}{\rho ^{1/2}}$. This gives
\begin{eqnarray*}
\delta V_q=-\frac{\hbar ^2}{4\mu\rho ^{1/2}}\frac{\partial ^2}{\partial x^2}(\frac{1}{\rho ^{1/2}}\delta \rho )+\frac{1}{4\mu\rho ^{3/2}}\frac{\partial ^2}{\partial x^2}\rho ^{1/2}\delta \rho \hspace{2.5cm} (A.11)
\end{eqnarray*}
By the integration by parts we simplify the 14-th term as follows
\begin{eqnarray*}
\int\int\lambda _2\rho\frac{\partial }{\partial x}\delta V_qdxdt&=&\int\lambda _2(x_r,t)\rho (x_r,t)\delta V_q(x_r,t)dt-\int\lambda _2(x_l,t)\rho (x_l,t)\delta V_q(x_l,t)dt\\\nonumber
&&+\int \frac{\partial}{\partial x}(\lambda _2\rho)\frac{1}{2\rho ^{1/2}}\frac{\partial}{\partial x}(\frac{1}{2\rho ^{1/2}}\delta \rho )|_{x_{l}}^{x_{r}}dt\\\nonumber
&&-\int\frac{\partial}{\partial x}(\frac{\partial}{\partial x}(\lambda _2\rho)\frac{1}{2\rho ^{1/2}})\frac{1}{2\rho ^{1/2}}|_{x_{r}}\delta \rho (x_r,t)dt \hspace{2.5cm} (A.12)\\\nonumber
&&+\int\frac{\partial}{\partial x}(\frac{\partial}{\partial x}(\lambda _2\rho)\frac{1}{2\rho ^{1/2}})\frac{1}{2\rho ^{1/2}}|_{x_{l}}\delta \rho (x_l,t)dt\\\nonumber
&&+\int\int\frac{\partial ^2}{\partial x^2}[\frac{\partial}{\partial x}(\lambda _2\rho)\frac{1}{2\rho ^{1/2}}]\frac{1}{2\rho ^{1/2}}\delta \rho (x,t)dxdt\\\nonumber
&&-\int\int\frac{\partial }{\partial x}(\lambda _2\rho )\frac{1}{4\rho ^{3/2}}\frac{\partial ^2}{\partial x^2}\rho ^{1/2}\delta \rho (x,t)dxdt
\end{eqnarray*}
~~~~~~~~~~~~Where $x_r$ and $x_l$ are the right and left ends of the one dimensional grid, and $F(x)|_{x_l}^{x_r}=F(x_r)-F(x_l)$ where $F(x)$ is any function. The first and the second terms in Eq.(A.12) are the contributions due to the change in the quantum potential at the two ends of the boundary only. Since, we take a large grid, $\rho $ at the two ends of the grid are very small and can be assumed constant. This leads to $V_q(x_r,t)$ and $V_q(x_l,t)$ being very high constant values at any time and hence $\delta V_q(x_r,t)=\delta V_q(x_l,t)=0$. With the same argument we can also neglect the contributions due to the terms 3rd, 4th and 5th. Combining all the terms we obtain the full variation in $\bar{J}$ as

\begin{eqnarray*}
\delta\bar{J}&=&\omega _a[<x>(T)-x_{cm}^d]\int x\delta \rho (x,T)dx-\int\int[\frac{\partial \rho}{\partial t}+\frac{\partial j}{\partial x}]\delta \lambda _1dxdt\\\nonumber
&&-\int\int[\frac{\partial j}{\partial t}+\frac{\partial}{\partial x}(\frac{j^2}{\rho}+\rho \frac{\partial}{\partial x}(V+V_q+V_{ext})\delta \lambda _2 dxdt\\\nonumber
&&-\int\lambda _1(x,T)\delta \rho (x,T)dx+\int\lambda _1(x,0)\delta \rho (x,0)dx \hspace{2.5cm} (A.13)\\\nonumber 
&&+\int\int\frac{\partial \lambda _1}{\partial t}\delta \rho dxdt-\int\lambda _2(x,T)\delta j(x,T)dx+\int\lambda _2(x,0)\delta j(x,0)dx\\\nonumber
&&+\int\int\frac{\partial \lambda _2}{\partial t}\delta j dxdt-\int\lambda _1(x_r,t)\delta j(x_r,t)dt+\int\lambda _1(x_l,t)\delta j(x_l,t)dt\\\nonumber
&&+\int\int\frac{\partial \lambda _1}{\partial x}\delta j(x,t)dxdt-2\int\frac{\lambda _2(x_r,t)j(x_r,t)}{\rho (x_r,t)}\delta j(x_r,t)dt\\\nonumber
&&+2\int\frac{\lambda _2(x_l,t)j(x_l,t)}{\rho (x_l,t)}\delta j(x_l,t)dt+\int\int\frac{\partial }{\partial x}(\lambda _2\frac{j}{\rho })\delta j(x,t)dxddt\\\nonumber
&&+\int\frac{\lambda _2(x_r,t)j^2(x_r,t)}{\rho ^2(x_r,t)}\delta \rho (x_r,t)dt-\int\frac{\lambda _2(x_r,t)j^2(x_r,t)}{\rho ^2(x_r,t)}\delta \rho (x_r,t)dt\\\nonumber
&&-\int\int\frac{\partial }{\partial x}(\lambda _2\frac{j^2}{\rho ^2})\delta \rho dxdt-2\int\int\lambda _2\frac{1}{\rho}\frac{\partial j}{\partial x}\delta jdxdt\\\nonumber
&&+2\int\int\lambda _2\frac{j}{\rho ^2}\frac{\partial \rho}{\partial x}\delta j dxdt+\int\int\lambda _2\frac{2j}{\rho ^2}\frac{\partial j}{\partial x}\delta \rho dxdt\\\nonumber
&&-2\int\int\lambda _2\frac{j^2}{\rho ^3}\frac{\partial \rho}{\partial x}\delta \rho dxdt-\int\int\lambda _2\frac{\partial}{\partial x}(V+V_q+V_{ext})\delta \rho dxdt\\\nonumber
&&-\frac{1}{\mu}\int\int\frac{\partial ^2}{\partial x^2}[\frac{\partial}{\partial x}(\lambda _2\rho)\frac{1}{2\rho ^{1/2}}]\frac{1}{2\rho ^{1/2}}\delta \rho dxdt\\\nonumber
&&+\frac{1}{\mu}\int\int\frac{\partial }{\partial x}(\lambda _2\rho)\frac{1}{4\rho ^{3/2}}\frac{\partial ^2}{\partial x^2}\rho ^{1/2}\delta \rho dxdt\\\nonumber
&&+\int\int\lambda _2\rho \frac{\partial }{\partial x}\mu (x)\delta E(t)dxdt
\end{eqnarray*}

~~~~~~~~~~~~This expression has 26 terms. Out of which, 5-th and 8-th terms can be dropped because we do not vary the initial density and quantum current. Again, 10-th, 11-th, 13-th, 14-th, 16-th and 17-th terms can also be dropped with the assumption that $\rho (x,t)$ and j(x,t) are very small at the boundary. Thus, the actual full variation in $\bar{J}$ becomes
\newpage
\begin{eqnarray*}
\delta\bar{J}&=&\omega _a[<x>(T)-x_{cm}^d]\int x\delta \rho (x,T)dx-\int\int[\frac{\partial \rho}{\partial t}+\frac{\partial j}{\partial x}]\delta \lambda _1dxdt\\\nonumber
&&-\int\int[\frac{\partial j}{\partial t}+\frac{\partial}{\partial x}(\frac{j^2}{\rho}+\rho \frac{\partial}{\partial x}(V+V_q+V_{ext})\delta \lambda _2 dxdt\\\nonumber
&&-\int\lambda _1(x,T)\delta \rho (x,T)dx+\int\int\frac{\partial \lambda _1}{\partial t}\delta \rho dxdt \hspace{2.5cm} (A.14)\\\nonumber
&&-\int\lambda _2(x,T)\delta j(x,T)dx+\int\int\frac{\partial \lambda _2}{\partial t}\delta j dxdt\\\nonumber
&&+\int\int\frac{\partial \lambda _1}{\partial x}\delta j(x,t)dxdt+\int\int\frac{\partial }{\partial x}(\lambda _2\frac{j}{\rho })\delta j(x,t)dxddt\\\nonumber
&&-\int\int\frac{\partial }{\partial x}(\lambda _2\frac{j^2}{\rho ^2})\delta \rho dxdt-2\int\int\lambda _2\frac{1}{\rho}\frac{\partial j}{\partial x}\delta jdxdt\\\nonumber
&&+2\int\int\lambda _2\frac{j}{\rho ^2}\frac{\partial \rho}{\partial x}\delta j dxdt+\int\int\lambda _2\frac{2j}{\rho ^2}\frac{\partial j}{\partial x}\delta \rho dxdt\\\nonumber
&&-2\int\int\lambda _2\frac{j^2}{\rho ^3}\frac{\partial \rho}{\partial x}\delta \rho dxdt-\int\int\lambda _2\frac{\partial}{\partial x}(V+V_q+V_{ext})\delta \rho dxdt\\\nonumber
&&-\frac{1}{\mu}\int\int\frac{\partial ^2}{\partial x^2}[\frac{\partial}{\partial x}(\lambda _2\rho)\frac{1}{2\rho ^{1/2}}]\frac{1}{2\rho ^{1/2}}\delta \rho dxdt\\\nonumber
&&+\frac{1}{\mu}\int\int\frac{\partial }{\partial x}(\lambda _2\rho)\frac{1}{4\rho ^{3/2}}\frac{\partial ^2}{\partial x^2}\rho ^{1/2}\delta \rho dxdt+\int\int\lambda _2\rho \frac{\partial }{\partial x}\mu (x)\delta E(t)dxdt
\end{eqnarray*}

\section*{References}
\begin{enumerate}
\item S. A. Rice, Science, {\bf 258}, 412 (1992)
\item D. J. Tannor and S. A. Rice, J. Chem. Phys. {\bf 83}, 5013 (1985)
\item A. P. Peire, M. A. Dahleh and H. Rabitz, Phys. Rev. {\bf A 37}, 4950 (1988)
\item D. J. Tannor, R. Kosloff and S. A. Rice, J. Chem. Phys., {\bf 85}, 5805 (1986)
\item R. Demiralp and H. Rabitz, Phys. Rev. {\bf A 47}, 809 (1993)
\item J. Botina and H. Rabitz, J. Chem. Phys. {\bf 104}, 4031 (1996)
\item S Shi and H. Rabitz, Comput. Phys. Comm., {\bf 63}, 71 (1991)
\item S Shi and H. Rabitz, Chem. Phys., {\bf 139}, 185 (1989)
\item W. Zhu, J. Botina and H. Rabitz, J. Chem. Phys., {\bf 108}, 1953 (1998)
\item Bijoy K. Dey, H. Rabitz and Attila Askar, Phys. Rev. {\bf A}, in press (2000)
\item D. J. Tannor and S. A. Rice, Adv. Chem. Phys., {\bf 70}, 441 (1988)
\item R. Kosloff, S. A. Rice, P. Gaspard, S. Tersigni and D. J. Tannor, Chem. Phys., {\bf 139}, 201 (1989)
\item P. Brumer and M. Shapiro, Faraday Discuss. Chem. Soc., {\bf 82}, 177 (1986)
\item P. Brumer and M. Shapiro, Annu. Rev. Phys. Chem., {\bf 43}, 257 (1992)
\item A. E. Bryson and Y. Ho, {\bf Applied Optimal Control} Hemisphere, New York (1975)
\item Bijoy K. Dey, Attila Askar and H. Rabitz, J. Chem. Phys., {\bf 109}, 8770 (1998)
\item Bijoy K. Dey, Attila Askar and H. Rabitz, Chem. Phys. Lett., {\bf 297}, 247 (1998)
\item D. Bohm, Phys. Rev., {\bf 85}, 166, 180 (1952)
\item D. Bohm, B. J. Hiley and P. N. Kaloyerou, Phys. Rep., {\bf 144}, 321 (1987)
\item K. P. Huber and G. Herzberg, {\bf Molecular Spectra and Molecular Structure IV. Constants of Diatomic Molecules}, Van Nostrand Reinhold co., NY, P.278 (1979)
\item J. P. Boris and D. L. Book, Methods in Comp. Phys., {\bf 16}, 85 (1976)
\item W. H. Press, B. P. Flannery, S. A. Teukolsky and W. T. Vetterling, {\bf Numerical Recipes}, Cambridge University, New York, (1992)
\item B. S. Rao, J. Phys. {\bf B4}, 791 (1971)
\item F. J. Belinfante, {\bf A Survey of Hidden-Variable Theories}, Pergamon press, Oxford, p. 188, (1973)
\end{enumerate}
\newpage
\section*{Figure Captions}
\begin{list}{\bfseries\upshape Figure \arabic{fig}:}
{\usecounter{fig}
\setlength{\labelwidth}{2cm}\setlength{\leftmargin}{2.6cm}
\setlength{\labelsep}{0.5cm}\setlength{\rightmargin}{1cm}
\setlength{\parsep}{0.5ex plus0.2ex minus 0.1ex}
\setlength{\itemsep}{0ex plus0.2ex} \slshape}
\item Optimal electric field shown as a function of time for $\Theta_T=3.0 a.u.$(solid line) and $\Theta _T=3.5 a.u.$(dotted line).
\item The expectation values $<x>$ shown as a function of time for $\Theta_T=3.0 a.u.$(solid line) and $\Theta _T=3.5 a.u.$(dotted line).
\item Initial(t=0)(dotted line) and final(t=T)(solid line) density corresponding to $\Theta_T=3.0 a.u.$(lebel a)) and $\Theta _T=3.5 a.u.$(label b)).
\item Hydrodynamical variables, viz., j(x,T)(b), $\lambda_1(x,T)$, $\lambda_2(x,T)$(c) and the real(solid) and imaginary(dotted) values of the wave function(a) plotted as a function of x. Notice that the hydrodynamical variables are smooth spatial function unlike the wave function.
\end{list}





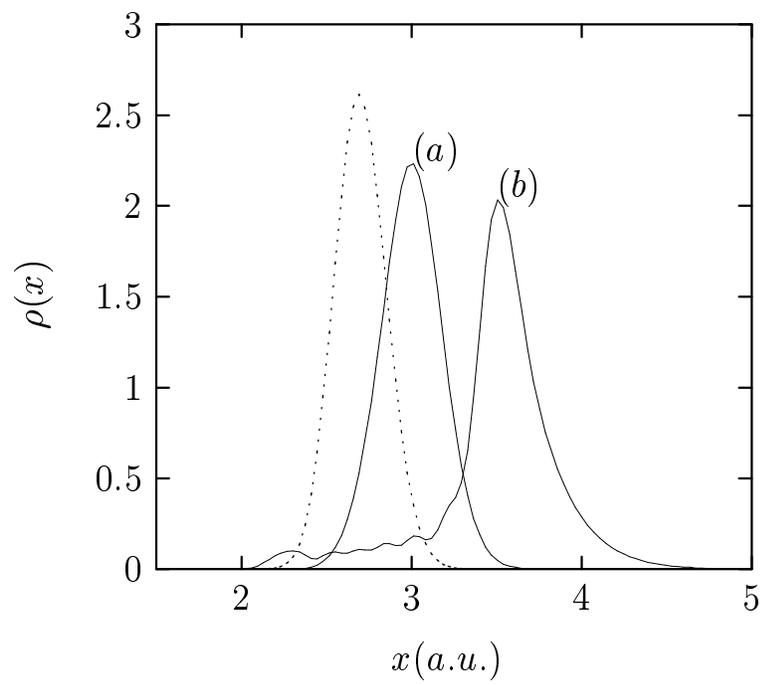
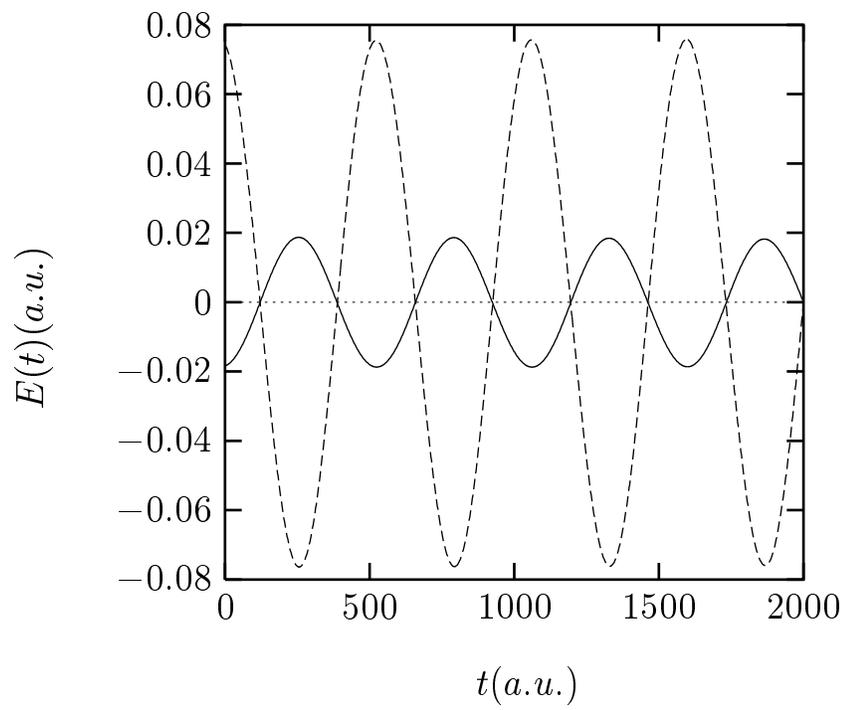
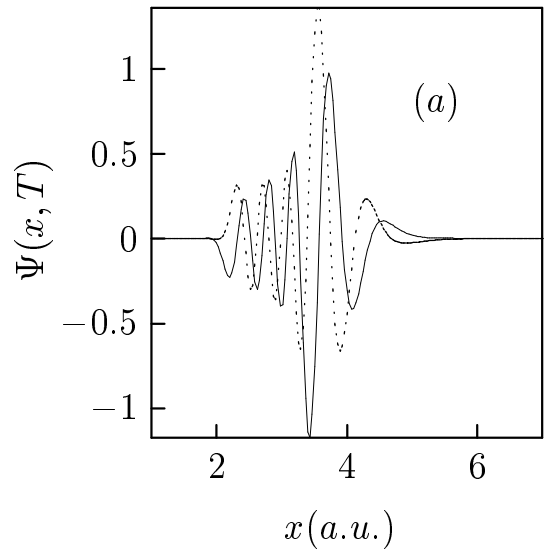
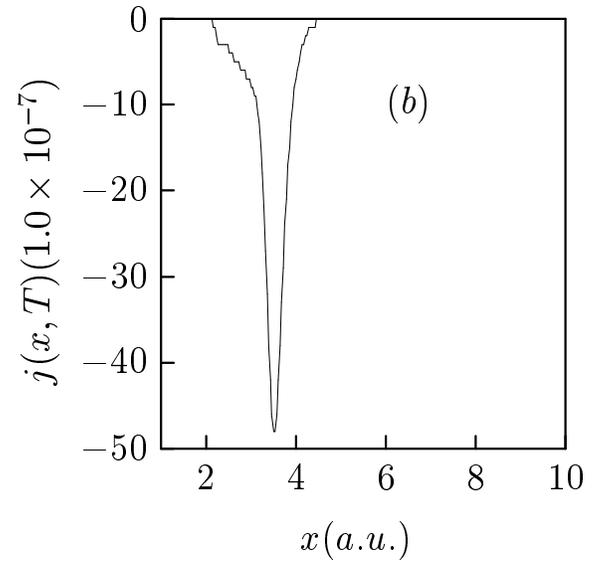
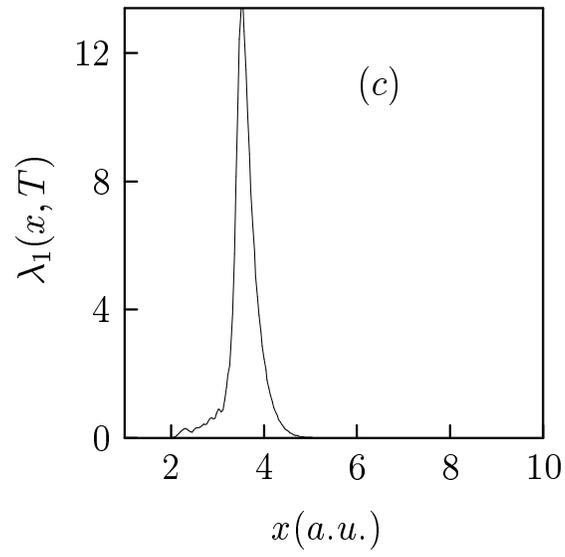
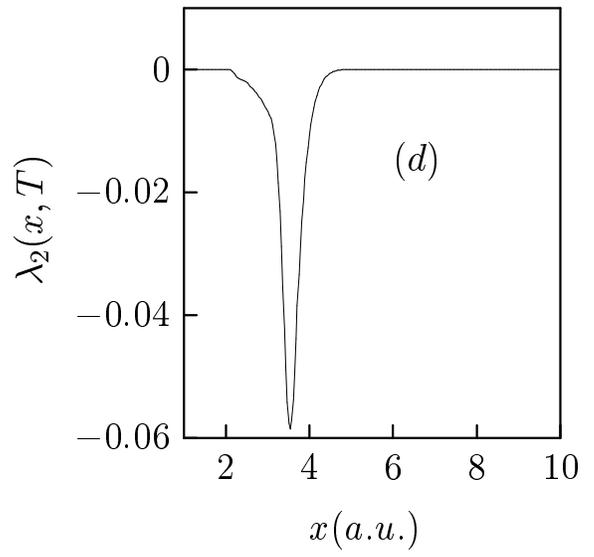
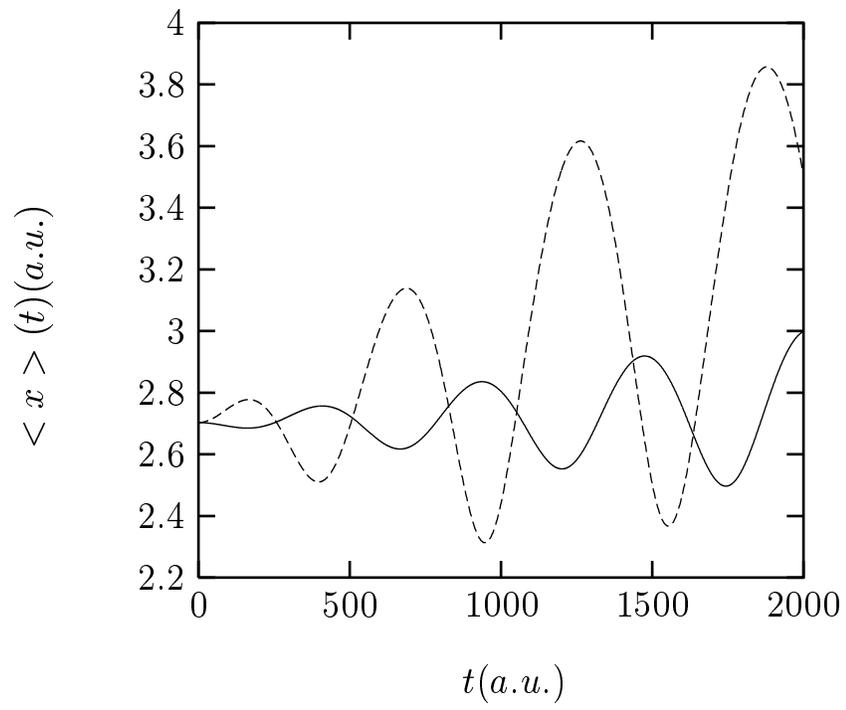
\end{document}